\documentclass[twocolumn,showpacs,amsmath,amssymb]{revtex4}
\usepackage{graphicx}% Include figure files
\usepackage{dcolumn}% Align table columns on decimal point
\usepackage{bm}% bold math
\usepackage{psfig}% bold math
\newcommand{\BR}{{\cal B}}
\newcommand{\pip}{\pi^+}
\newcommand{\pim}{\pi^-}
\newcommand{\piz}{\pi^0}
\newcommand{\psp}{\psi^{\prime}}
\newcommand{\psip}{\psi(2S)}
\newcommand{\pspp}{\psi^{\prime \prime}}
\newcommand{\psipp}{\psi(3770)}
\newcommand{\jpsi}{J/\psi}
\newcommand{\EE}{e^+e^-}
\newcommand{\MM}{\mu^+\mu^-}
\newcommand{\TT}{\tau^+\tau^-}
\newcommand{\PP}{\pi^+\pi^-}
\newcommand{\KK}{K^+K^-}
\newcommand{\kskl}{K^0_SK^0_L}
\newcommand{\OP}{\omega\pi^0}
\newcommand{\RP}{\rho\pi}
\newcommand{\rhopi}{\rho\pi}
\newcommand{\ccb}{c\overline{c}}
\newcommand{\BBb}{B\overline{B}}
\newcommand{\ppb}{p\overline{p}}
\newcommand{\LLb}{\Lambda \overline{\Lambda}}
\newcommand{\ddb}{D\overline{D}}

\newcommand{\PPJP}{\pi^+\pi^- J/\psi}
\newcommand{\ra}{\rightarrow}
\newcommand{\EETO}{e^+e^-\to}
\newcommand{\beq}{\begin{equation}}
\newcommand{\eeq}{\end{equation}}
\newcommand{\beqn}{\begin{eqnarray}}
\newcommand{\eeqn}{\end{eqnarray}}
\newcommand{\bnum}{\begin{enumerate}}
\newcommand{\enum}{\end{enumerate}}
\newcommand{\btbl}{\begin{table}}
\newcommand{\etbl}{\end{table}}
\newcommand{\btbu}{\begin{tabular}}
\newcommand{\etbu}{\end{tabular}}
\begin{document}
\preprint{Draft-PRD}

\title{\boldmath Possible large branching fraction of $\pspp$ decays 
to charmless final states}
\author{P.~Wang}
 \email{wangp@mail.ihep.ac.cn}
\author{C.~Z.~Yuan}
\author{X.~H.~Mo}
\affiliation{Institute of High Energy Physics, CAS, Beijing
100039, China}

\date{\today}

\begin{abstract}
The branching fraction of the $\pspp$ decays into charmless final
states is estimated in the $S$- and $D$-wave charmonia mixing
scheme. With the information of  the hadronic decays of $\jpsi$ and $\psp$, it is found that a large branching fraction, up to 13\% of 
the total $\pspp$ decays may go to charmless final states. The
experimental search for these decays is also discussed.
\end{abstract}


\maketitle

\section{Introduction}

As more and more data are collected at $\psipp$ (shortened as
$\pspp$), 27.7~pb$^{-1}$ by BES-II, 55~pb$^{-1}$ by CLEOc, and a
few fb$^{-1}$ is in progress by CLEOc, the study of the non-$\ddb$
decays of $\pspp$ gets renaissance recently.

Due to its mass coincides with the $\ddb$ mass threshold, $\pspp$
is believed to decay predominantly to $\ddb$, and the non-$\ddb$
decays including transitions between charmonium states and decays
into light hadrons are expected to be small. The first exclusive
non-$\ddb$ decay mode of $\pspp$ was searched for using the
Mark-III data sample~\cite{zhuyn}, where evidence of
$\pspp\ra\jpsi\pip\pim$ was observed; afterwards theoretical
calculation of its decay width was published~\cite{kuang}. 
This decay mode was searched for recently by BES and a signal of
3.1$\sigma$ was observed~\cite{besppjp}, which is in marginally
agreement with the upper limit set by CLEOc~\cite{cleocppjp}.  
The other searches of the non-$\ddb$ decays of $\pspp$ were all 
reported in either doctoral theses~\cite{zhuyn,majid} based on 
the Mark-III data or conference talk~\cite{ichep04_rongg} based on 
the BES-II data, and no statistically significant results were given. This may indicate that the data samples utilized are still not large 
enough to search for these channels of small branching fractions.
Now with much larger $\pspp$ samples, measurements of the cross 
section $\sigma(\EETO \pspp \ra \ddb) \equiv \sigma(\ddb)$ have 
recently been reported by BES-II~\cite{ichep04_rongg} and
CLEOc~\cite{cleoddb} collaborations using either single-tag or 
double-tag method. Comparing with the measured
$\sigma(\EETO \pspp \ra \mbox{anything})$, the difference can be up to 
1.4~nb (about 18\% of the total cross section of the $\pspp$
production in $\EE$ collision~\cite{rosner}), which indicates the 
existence of substantial non-$\ddb$ decays for $\pspp$. However, there is limited theoretical work on this aspect. In Ref.~\cite{rosner}, it is
estimated that at most 600~keV ($\sim 2.5$\%) of the $\pspp$ total width 
of $(23.6 \pm 2.7)$~MeV is due to the radiative transition, and perhaps 
as much as another 100~keV ($\sim 0.4$\%) is due to the hadronic 
transition to $\jpsi\pi\pi$. All these together are far from accounting 
for a deficit of 18\% of the total $\pspp$ width.

In this paper, we concentrate on the charmless decays of
$\pspp$. By charmless decay, we exclude those decay modes with
either open or hidden charm. We shall estimate the partial width
of the charmless decay of $\pspp$ in the $2S$-$1D$ charmonia
mixing scenario, under the assumption that the perturbative QCD
(pQCD) ``12\% rule'' holds between pure $1S$ and $2S$ charmonium
states. The mixing between $2S$ and $1D$ states of charmonium was
originally proposed to explain the measured large $\Gamma_{ee}$ of
$\pspp$. It is suggested~\cite{eichten2} that the mass eigenstates 
$\psi(3686)$ (shortened as $\psp$) and $\pspp$ are the mixtures of 
the $2S$- and $1D$-wave of charmonia, namely $\psi(2^3S_1)$ and 
$\psi(1^3D_1)$ states. This was used by Rosner~\cite{rosnersd} in 
explaining the ``$\rho\pi$ puzzle'' in $\psp$ and $\jpsi$ decays. 
He suggested that the mixing of $\psi(2^3 S_1)$ and $\psi(1^3 D_1)$ 
states is in such a way which leads to almost complete cancellation 
of the decay amplitude of $\psp \rightarrow \rhopi$, and the missing 
$\rhopi$ decay mode of $\psp$ shows up instead as enhanced decay mode of $\pspp$. This idea was then applied to solve the enhanced decay of $\psp \to \kskl$ relative to $\jpsi \ra \kskl$, and a prediction of the
branching fraction of $\pspp\rightarrow \kskl$ was 
given~\cite{psippkskl}. In principle, if this scenario is correct, it can be generalized and relate the partial widths of each individual 
mode in $\jpsi$, $\psp$ and $\pspp$ decays. 
Therefore by virtue of the branching fractions of $\jpsi$ and
$\psp$ decays, we are able to estimate the corresponding decay 
branching fraction of $\pspp$. This will be tested by the upcoming 
large $\pspp$ data sample from CLEOc.

In the following parts of the paper, we begin our study with a
general review of the 12\% rule between $\psp$ and $\jpsi$ decays,
followed by an introduction of the $2S$- and $1D$-wave charmonia
mixing scheme. Then we conduct the calculation on the possible
charmless decay width of $\pspp$ with the available information
from $\jpsi$ and $\psp$ decays.  We show that the scenario and
currently available information on $\jpsi$ and $\psp$ decays can
accommodate a partial width of 3.0~MeV charmless decays of
$\pspp$. Finally we turn to the experimental searches for these
decays. Both inclusive and exclusive methods are examined.

\section{12\% rule and $2S$-$1D$ mixing{\label{II}}}

From the pQCD, it is expected that both $\jpsi$ and $\psp$
decaying into light hadrons are dominated by the annihilation of
$c\bar{c}$ into three gluons, with widths proportional to the
square of the wave function at the origin
$|\Psi(0)|^2$~\cite{appelquist}. This yields the pQCD 12\% rule,
that is
\begin{eqnarray}
Q_h &=&\frac{{\cal B}_{\psp \ra h}}{{\cal B}_{\jpsi \ra h}}
=\frac{{\cal B}_{\psp \ra \EE}}{{\cal B}_{\jpsi \ra \EE}} 
\approx 12.7\%. \label{qcdrule}
\end{eqnarray}

The violation of the above rule was first observed in $\rhopi$ and
$K^{*+}K^-+c.c.$ modes by Mark-II~\cite{mk2}. Since then BES-I,
BES-II and CLEOc have measured many two-body decay modes of
$\psp$~\cite{bes1vt,bes2vt,jpsikskl,psipkskl,besksk,besrop,besvpn,besogp,cleocvp}. Among them, some obey the rule , like baryon-antibaryon ($\BBb$) modes, while others are either suppressed in $\psp$ decays, like 
vector-pseudoscalar (VP) and vector-tensor (VT) modes, or enhanced, 
like $\kskl$. There have been many theoretical efforts trying to solve the puzzle~\cite{puzzletheory,rosnersd}. Among them, the $2S$-$1D$
charmonia mixing scenario~\cite{rosnersd} predicts with little
uncertainty $\BR(\pspp \ra \rhopi)$ which agrees with experimental
data~\cite{wympspp,zhuyn}.

In $S$- and $D$-wave charmonia mixing scheme, the mass eigenstates
$\psp$ and $\pspp$ are the mixtures of the $S$- and $D$-wave
charmonia, namely 
\begin{equation*}
\begin{array}{l}
 |\psp\rangle = | 2^3 S_1 \rangle \cos \theta
                  - | 1^3 D_1 \rangle \sin \theta~, \\
 |\pspp\rangle = | 2^3 S_1 \rangle \sin \theta
                  + | 1^3 D_1 \rangle \cos \theta~,
\end{array}
%\label{sdmix} 
\end{equation*} 
where $\theta$ is the mixing angle between pure
$\psi(2^3 S_1)$ and $\psi(1^3D_1)$ states and is fitted from the
leptonic widths of $\pspp$ and $\psp$ to be either $(-27 \pm
2)^{\circ}$ or $(12 \pm 2)^{\circ}$~\cite{rosnersd}. The latter
value of $\theta$ is consistent with the coupled channel
estimates~\cite{eichten2,heikkila} as well as the ratio between $\psp$
and $\pspp$ partial widths to $\jpsi\pip\pim$~\cite{kuang}.
Hereafter, the calculations and discussions in this paper are
solely for the mixing angle $\theta =12^{\circ}$~\cite{ktchao}.

As in the discussion of Ref.~\cite{rosnersd}, since both hadronic
and leptonic decay rates are proportional to the square of the
wave function at the origin, it is expected that if $\psp$ is a
pure $\psi(2^3 S_1)$ state, then for any hadronic final states
$f$,
 \beq
 \Gamma(\psp \ra f)=\Gamma(\jpsi \ra f) \frac{\Gamma(\psp \ra \EE)}
 {\Gamma(\jpsi \ra \EE)}~. \label{asu}
 \eeq

The electronic partial width of $\jpsi$ is expressed in potential
model by~\cite{novikov} \begin{equation*} \Gamma(\jpsi \ra \EE)
=\frac{4\alpha^2 e^2_c}{M_{\jpsi}^2} \left|R_{1S}(0)\right|^2,
\label{jee} \end{equation*} with $\alpha$ the QED fine structure constant,
$e_c=2/3$, $M_{\jpsi}$ the $\jpsi$ mass and $R_{1S}(0)$ the radial
$1^3S_1$ wave function at the origin.

Since $\psp$ is not a pure $\psi(2^3 S_1)$ state, its electronic
partial width is expressed as~\cite{rosnersd} 
\begin{eqnarray*} \Gamma(\psp
\ra \EE) & =& \frac{4\alpha^2 e^2_c}{M_{\psp}^2}
%\label{sipee} 
\\
&\times& \left|\cos \theta R_{2S}(0) - \frac{5}{2\sqrt{2}m_c^2}
\sin \theta R^{\prime\prime}_{1D}(0) \right|^2, % \nonumber \eeqn
\end{eqnarray*}
with $M_{\psp}$ the $\psp$ mass, $m_c$ the $c$-quark mass,
$R_{2S}(0)$ the radial $2^3S_1$ wave function at the origin and
$R^{\prime\prime}_{1D}(0)$ the second derivative of the radial
$1^3D_1$ wave function at the origin. In the calculations in this
paper, we take $R_{2S}(0)=0.734$~GeV$^{3/2}$ and
$5R^{\prime\prime}_{1D}(0)/(2\sqrt{2}m_c^2)=0.095$~GeV$^{3/2}$ from
Refs.~\cite{rosnersd, rosnerm}.

If Eq.~(\ref{asu}) holds for a pure $2S$ state, $\pspp \ra f$,
$\psp \ra f$ and $\jpsi \ra f$ partial widths are to
be~\cite{psippkskl} 
\beqn
\Gamma(\pspp \ra f) & = & \frac{C_f}{M_{\pspp}^2} \left| \sin
\theta R_{2S}(0) + \eta \cos \theta \right|^2,
     \nonumber \\
\Gamma(\psp \ra f) & = & \frac{C_f}{M_{\psp}^2} \left| \cos \theta
R_{2S}(0) - \eta \sin \theta \right|^2,
       \nonumber \\
\Gamma(\jpsi \ra f) & = & \frac{C_f}{M_{\jpsi}^2} \left|R_{1S}(0)
\right|^2~, \label{tof} \eeqn where $C_f$ is a common factor for
the final state $f$, $M_{\pspp}$ the $\pspp$ mass, and
$\eta=|\eta| e^{i\phi}$ is a complex parameter with $\phi$ being
the relative phase between $\langle f|1^3D_1 \rangle$ and $\langle
f|2^3S_1 \rangle$.

From Eq.~(\ref{tof}), it is obvious that with $\Gamma(\jpsi
\ra f)$ and $\Gamma(\psp \ra f)$ known, two of the three parameters,
$C_f$ and complex $\eta$, can be fixed, thus can be used to predict
$\Gamma(\pspp \ra f)$ with only one unknown parameter, say, the
phase of $\eta$. %Since $\eta$ is proportional to 
%$\langle f|1^3D_1 \rangle / \langle f|2^3S_1 \rangle$, so its phase 
%is the  relative phase between $\langle f|1^3D_1 \rangle$ and 
%$\langle f|2^3S_1 \rangle$. 
Thus the $S$- and $D$-wave mixing scenario provides
a mathematical scheme to calculate the partial
width of $\pspp$ decay to any exclusive final state,  
with its measured partial widths in $\jpsi$ and $\psp$
decays.

However, the current information concerning the $\psp$ decay is
extremely limited, which prevents us from estimating $Q_h$ values
for most exclusive decay modes. Table~\ref{qval} lists some hadronic 
final states which are measured both in $\jpsi$ and $\psp$ decays, 
together with the calculated $Q_h$ defined in Eq.~(\ref{qcdrule}). 
Summing up all the channels in Table~\ref{qval} makes less than 2\%
of the $\psp$ decay through $ggg$ annihilation.

From Table~\ref{qval}, we notice that compared with the 12\% rule,
the $\psp$ decays to
 \bnum \item the pseudoscalar-pseudoscalar (PP) mode $\kskl$
is enhanced; \item the VP and VT modes are suppressed; \item most
of the $\BBb$ modes are consistent with it.
 \enum
The summed branching fractions and $Q_h$ values for these three
categories of decay modes are evaluated and also listed in 
Table~\ref{tyqval}. In estimating the charmless decays of $\pspp$, we 
shall discuss these three different cases separately.

\begin{table*}[htbp]
\caption{\label{tyqval}Branching fractions and $Q_h$ values for
some $\jpsi$ and $\psip$ decay channels. } \center
\begin{ruledtabular}
{\small %\footnotesize
\begin{tabular}{cccccc}
  Modes  &  Channels
                &${\cal B}_{\jpsi} (10^{-3})$
                                  &${\cal B}_{\psip} (10^{-4})$
                                                  &$Q_h$ (\%)
                          & Ref. \\ \hline
$0^-0^-$ &$\PP$  &$0.147\pm0.023$ &$0.8 \pm 0.5 $ & $ 54 \pm 35  $
                                                  &\cite{pdg} \\
         &$\KK$  &$0.237\pm0.031$ &$1.0 \pm 0.7 $ & $ 42 \pm 30  $
                                              & \cite{pdg}  \\
         &$\kskl$&$0.182\pm0.014$ &$0.52\pm 0.07$ &$28.8\pm 3.7$
                                &\cite{jpsikskl,psipkskl} \\ \hline
sum      & PP    &$0.57 \pm 0.07$ &$2.32\pm 1.27$ &$41.1\pm 22.8 $
                                                  &      \\ \hline
$1^-0^-$ &$\RP$  & $12.7\pm 0.9 $ &$0.29\pm 0.07$ & $0.23\pm 0.6 $
                                                  &\cite{resvp} \\
         &$K^+ \overline{K}^* (892)^- +c.c.$
                 & $5.0 \pm 0.4 $ &$0.15\pm 0.08$ & $0.3\pm 0.2$
                                                  & \\
         &$K^0 \overline{K}^* (892)^0 +c.c.$
             & $4.2 \pm 0.4 $ &$1.10\pm 0.20$ & $2.6 \pm 0.5$ &$   $ \\
         &$\OP$  & $0.42\pm 0.06$ &$0.20\pm 0.06$ & $4.8 \pm 1.6$
                                           &   \\ \hline
$1^-2^+$ &$\omega f_2(1270)$
                 & $ 4.3\pm 0.6 $ &$2.05\pm 0.56$ & $4.8 \pm 1.5$
                                                  &\cite{bes2vt} \\
         &$\rho a_2 $
                 & $10.9\pm 2.2 $ &$2.55\pm 0.87$ & $2.3 \pm 1.0$
                                          &$   $ \\
         &$K^*(892)^0  \overline{K}_2^* (1430)^0 +c.c.$
                 & $ 6.7\pm 2.6 $ &$1.86\pm 0.54$ & $2.8 \pm 1.3$
                                          &$   $ \\
         &$\phi f_2^{\prime} (1525)$
                 & $1.23\pm 0.21$ &$0.44\pm 0.16$ & $3.6 \pm 1.4$
                                          &$   $ \\ \hline
 sum     &VP\& VT& $45.45\pm 7.37$&$8.64\pm 2.54$ & $1.90\pm 0.64$
                                                  &\\ \hline
$\BBb$
         &$\ppb$ & $2.12\pm 0.10$ &$2.07\pm 0.31$ & $9.8 \pm 1.5$
                                                  &\cite{pdg} \\
         &$\LLb$ & $1.30\pm 0.12$ &$1.81\pm 0.34$ & $13.9\pm 2.9$
                                              &$   $ \\
         &$\Sigma^0 \overline{\Sigma}^0$
                 & $1.27\pm 0.17$ &$1.2 \pm 0.6 $ & $9.4 \pm 4.9$
                                          &$   $ \\
%        &$\Sigma^{*+} \overline{\Sigma}^{*-}$
         &$\Sigma(1385)^{\pm} \overline{\Sigma}(1385)^{\mp}$
                 & $1.03\pm 0.13$ &$1.1 \pm 0.4 $ & $10.7\pm 4.1$ &$   $ \\
         &$\Xi \overline{\Xi}$
                 & $1.8 \pm 0.4 $ &$1.88\pm 0.62\dagger$
                                          & $10.4\pm 4.2$ &$   $ \\
%        &$\Xi^- \overline{\Xi}^+$
%                 & $ - $          &$0.94\pm 0.31$ & $ - $         &$   $ \\
         &$\Delta^{++} \overline{\Delta}^{--}$
                 & $1.10\pm 0.29$ &$1.28\pm 0.35$ & $11.6\pm 4.4$ &$   $ \\
%         &$\Xi^{*0} \overline{\Xi}^{*0}$
%                 & $ - $          &$<0.81   $     & $ - $         &$   $ \\
                                                         \hline
 sum     &$\BBb$ & $8.62\pm 1.21$ &$9.34\pm 2.62$ & $10.8\pm 3.4$ &\\
                                                          %% \hline\hline
\end{tabular} \\
Note: $\dagger$ simple normalization by $\Xi^- \overline{\Xi}^+ =
(1/2) \Xi \overline{\Xi}$. } \label{qval}
\end{ruledtabular}
\end{table*}

Since the experimental information on the exclusive decays of
$\psp$ is rather limited, we turn to inclusive branching fractions of 
$\jpsi$ and $\psp$ hadronic decays as an alternative. The estimation is 
based on the assumption that the decays of $\jpsi$ and $\psp$ in the 
lowest order of QCD are classified into hadronic decays ($ggg$), 
electromagnetic decays ($\gamma^*$), radiative decays into light hadrons 
($\gamma gg$), and transition to lower mass charmonium states 
($\ccb X$)~\cite{kopke,guli}. Thus, using the relation 
$\BR(ggg)+\BR(\gamma gg)+\BR(\gamma^*)+\BR(\ccb X)=1$, one can derive
$\BR(ggg)+\BR(\gamma gg)$ by subtracting $\BR(\gamma^*)$ and
$\BR(\ccb X)$ from unity.

The calculated values of $\BR(\gamma^*)$ and $\BR(\ccb X)$, together with the values used to calculate them are summarized in Table~\ref{esdat}. As regards to $\psp$, two final states $\gamma \eta (2S)$ and 
$h_c(1^1P_1)+X$ with faint branching fractions are neglected in our
calculation. By deducting the contributions $\BR(\gamma^*)$ and 
$\BR(\ccb X)$, we find that $\BR(\jpsi \ra ggg)+\BR(\jpsi \ra \gamma gg)= (73.5\pm 0.6)\%$ and $\BR(\psp \ra ggg)+\BR(\psp \ra \gamma gg)=
(19.1\pm2.5)\%$, then the ratio of them is 
%Therefore the ratio of the branching fractions of $\psp$ to $\jpsi$ decays into hadrons is 
 \beq Q_g =\frac{\BR(\psp \ra ggg + \gamma gg)}
               {\BR(\jpsi \ra ggg + \gamma gg)} = (26.0 \pm 3.5)\%~.
 \label{qvone} \eeq
The above estimation is consistent with the previous
ones~\cite{suzuki,guli}. The relation between the decay rates of
$ggg$ and $\gamma gg$ is readily calculated in pQCD to the first
order as~\cite{kwong}
 \begin{equation*} 
\frac{\Gamma(\jpsi \ra \gamma gg)}{\Gamma(\jpsi \ra ggg)}
 = \frac{16}{5} \frac{\alpha}{\alpha_s(m_c)}
 \left( 1- 2.9 \frac{\alpha_s}{\pi} \right). %\label{rggg} 
\end{equation*}
Using $\alpha_s(m_c)=0.28$, one can estimate the ratio to be
0.062. A similar relation can be deduced for the $\psp$ decays.
So we obtain $\BR(\jpsi \ra ggg) \simeq (69.2\pm0.6)\%$ and
$\BR(\psp \ra ggg) \simeq (18.0\pm 2.4)\%$, while the ``26.0\%
ratio'' in Eq.~(\ref{qvone}) stands well for both $ggg$ and
$\gamma gg$. Although $Q_g$ is considerably enhanced relative to
$Q_h$ in Eq.~(\ref{qcdrule}), it coincides with the ratio for the
$\kskl$ decay mode between $\psp$ and $\jpsi$, which is \beq Q_{\kskl}
=(28.8 \pm 3.7)\%~, \label{qvkskl} \eeq according to the recent results
from BES~\cite{jpsikskl,psipkskl}. The relation in
Eq.~(\ref{qvone}) was discussed in the literature as the hadronic
excess in $\psp$ decay~\cite{guli,suzuki}. It implicates that
while some modes are suppressed in $\psp$ decays, the dominant part of
$\psp$ through $ggg$ decays is enhanced relative to the 12\% rule 
prediction in the light of $\jpsi$ decays.

 \btbl[htb]
\caption{\label{esdat}Experimental data on the branching fractions
for $\jpsi$ and $\psp$ decays through virtual photon and to lower
mass charmonium states used in this analysis. Most of the data are
taken from PDG~\cite{pdg}, except for $\BR (\jpsi,\psp \ra \gamma^* \ra \mbox{hadrons})$, which are calculated by the product $R
\cdot \BR (\jpsi,\psp \ra \MM)$, with $R=2.28\pm 0.04$~\cite{seth}. 
In estimating the errors of the sums, the correlations between the 
channels are considered.} \vskip 0.2 cm \center
\btbu{lcc}
\hline \hline Channel & $\BR(\jpsi)$ & $\BR(\psp) $        \\
\hline \hline $\gamma^* \ra \mbox{hadrons}$
        & (13.4$\pm$0.33)\%  & (1.66$\pm$0.18)\%  \\
$\EE$   & (5.93$\pm$0.10)\% & (7.55$\pm$0.31)$\times 10^{-3}$ \\
$\MM$   & (5.88$\pm$0.10)\% &  (7.3 $\pm$0.8)$\times 10^{-3}$ \\
$\TT$   &                   &  (2.8 $\pm$0.7)$\times 10^{-3}$ \\
\hline $\gamma^* \ra X$
        &(25.22$\pm$0.43)\% & (3.43$\pm$0.27)\%  \\ \hline
$\gamma\eta_c$
        & (1.3$\pm$0.4)\%   & (2.8$\pm$0.6)$\times 10^{-3}$ \\
$\PPJP$ &                   & (31.7$\pm$1.1)\%   \\
$\piz\piz\jpsi$ &           & (18.8$\pm$1.2)\%   \\
$\eta\jpsi$     &           & (3.16$\pm$0.22)\%  \\
$\piz\jpsi$     &           & (9.6$\pm$2.1)$\times 10^{-4}$ \\
$\gamma\chi_{c0}$&          & (8.6$\pm$0.7)\%    \\
$\gamma\chi_{c1}$&          & (8.4$\pm$0.8)\%    \\
$\gamma\chi_{c2}$&          & (6.4$\pm$0.6)\%    \\  \hline $\ccb
X$& (1.3$\pm$0.4)\%   &(77.4$\pm$2.5)\%    \\ \hline \hline \etbu
\etbl

%It should be noted that the improvement of the $\psp$ transition
%branching fractions may affect the estimation of $Q_g$, thus affect the 
%evaluation of $\pspp$ charmless decay branching fraction in the next
%section. However, comparing with current uncertainty of experiments, 
%these changes are out of consideration in this paper.

\section{The charmless decays of $\pspp${\label{III}}}

We define the enhancement or suppression factor as~\cite{rosner}
\beq
Q(f) \equiv \frac{\Gamma(\psp \ra f) }{\Gamma(\jpsi \ra f)}
         \frac{\Gamma(\jpsi \ra \EE)}{\Gamma(\psp \ra \EE)}.
\label{definec} \eeq
In the $2S$-$1D$ mixing scheme, for any final state $f$, its partial
width in 
$\pspp$ decay can be related to its partial width in $\jpsi$ and
$\psp$ decays by Eq.(\ref{tof}),
with an unknown parameter $\phi$ which is the phase of $\eta$. This
unknown 
phase constrains the predicted $\Gamma(\pspp \ra f)$ in a finite range.
We calculate
\beq
R_\Gamma \equiv \Gamma(\pspp \ra f) / \Gamma(\jpsi \ra f)
\label{rgamma}
\eeq
as a function of $Q(f)$ and plot it in Fig.~\ref{enhsup}. In the
figure, the solid contour corresponds to the solution with
$\phi=0$; the dashed one corresponds to the solution with
$\phi=180^\circ$; and the hatched area corresponds to the
solutions with $\phi$ taking other non-zero values.

To make it clear, we discuss the final states in three situations:
$Q(f)<1$, $Q(f)>1$, and $Q(f) = 1$.

\subsection{Final states with $Q(f)<1$}

If $Q(f) <1$, the decay $\psp \ra f$ is suppressed relative to
$\jpsi \ra f$. The extreme situation is $Q(f) \rightarrow 0$,
corresponding to the absence of the mode $f$ in $\psp$ decays.
This is the case which was assumed for the $\rhopi$ mode in the
original work to solve the $\rhopi$ puzzle by the $S$- and
$D$-wave mixing~\cite{rosnersd}. If $Q(f)=0$, the solution of the
second equality of Eq.~(\ref{tof}) simply yields $\eta = R_{2S}(0)
\cos \theta / \sin \theta$ which cannot have a non-zero phase, and
$R_\Gamma=9.2$.
%$R_\Gamma=9.16$.

\begin{figure}[htbp]
\includegraphics[width=7cm]{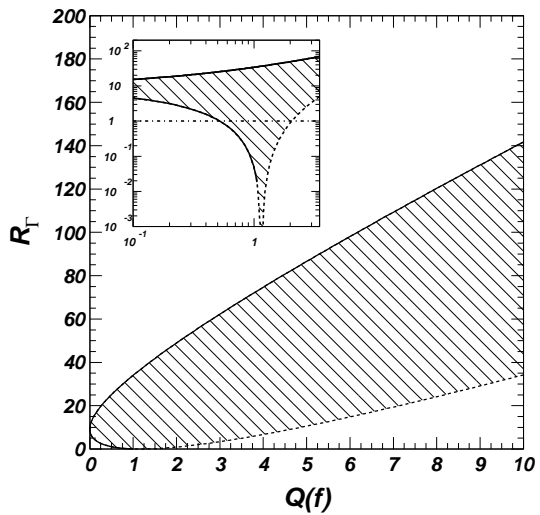}
\caption{\label{enhsup} $R_{\Gamma}$ versus $Q(f)$. The solid
contour corresponds to $\phi=0$; the dashed contour corresponds to
$\phi=180^\circ$; and the hatched area corresponds to $\phi$
having other non-zero values. The inset displays the variation of
$R_{\Gamma}$ in the vicinity of $Q(f)=1$, where the dot-dashed
line denotes $R_{\Gamma}=1$.}
\end{figure}

Generally, the suppression factor could be different from zero.
Even the $\rhopi$ and other strongly suppressed VP modes are found
in $\psp$ decays recently by BESII~\cite{besrop} and
CLEOc~\cite{cleocvp} with $Q(VP) \sim {\cal O} (10^{-2})$. In this case,
there are two real and positive solutions of $\eta$ as shown in
Fig.~\ref{enhsup} corresponding to the maximum and minimum of
their possible partial widths in $\pspp$ decays. The solutions
with $\eta$ having a non-zero phase yield the values of $R_\Gamma$
between the minimum and maximum limits.

For VT final states, which are measured to have $Q(VT) \approx
1/3$, with Eq.~(\ref{tof}), we get $2.0 \le R_\Gamma \le 21.6$, as
shown in Fig.~\ref{enhsup}, where the upper and lower limits
correspond to two real and positive solutions of $\eta$, the range
is due to the values of $\eta$ with non-zero phases.

\subsection{Final states with $Q(f)>1$}

If $Q(f) > 1$, the decay $\psp \ra f$ is enhanced relative to
$\jpsi \ra f$. The extreme situation is $Q(f) \rightarrow \infty$,
corresponding to the complete absence of the final state $f$ in $\jpsi$
decays. For
$$Q(f) > \left| \frac{ \cos\theta R_{2S}(0)}
{\cos\theta R_{2S}(0) - \frac{5}{2\sqrt{2} m_c^2} \sin\theta
  R_{1D}^{\prime\prime}(0) } \right|^2 = 1.06,$$
there are two real solutions of $\eta$, one is positive and the
other negative. From the first equality of Eq.~(\ref{tof}), it is
seen that the positive solution leads to the larger $R_\Gamma$.
i.e. larger $\Gamma (\pspp \ra f)$ (the solid contour in
Fig.~\ref{enhsup}) while the negative solution leads to the
smaller $\Gamma(\pspp \ra f)$ (the dashed contour in
Fig.~\ref{enhsup}).

For the known enhanced mode in $\psp$ decays like $\kskl$,
$Q(\kskl) = 2.26$, we find $1.4 \le R_\Gamma \le 52.5$, which
corresponds to the $\pspp$ decay partial width from 0.024 to
0.87~keV.

\begin{figure}
\includegraphics[width=7cm]{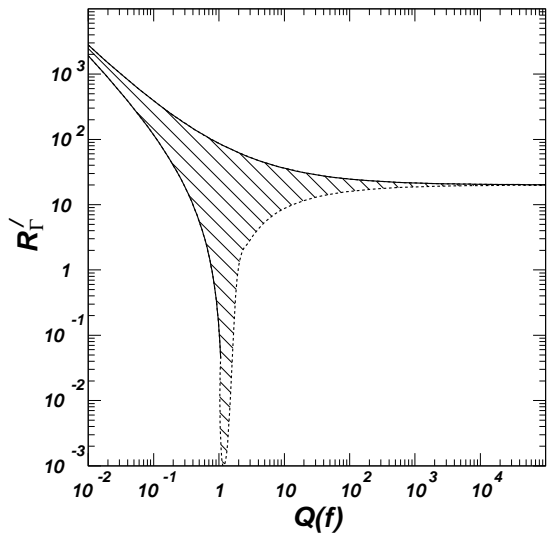}
\caption{\label{enh} $R^\prime_{\Gamma}$ versus $Q(f)$. The solid
contour corresponds to $\phi=0$; dashed contour corresponds to
$\phi=180^\circ$; and the hatched area corresponds to $\phi$
having other non-zero values.}
\end{figure}

It should be noted that for finite $\Gamma(\psp \ra f)$, $Q(f)
\rightarrow \infty$ means the diminishing of $\Gamma(\jpsi \ra
f)$, which gives rise to $R_\Gamma \rightarrow \infty$ according
to the definition of $R_\Gamma$ in Eq.(\ref{rgamma}). 
Under such circumstance, it is more
intuitive to calculate
 \[  R^\prime_\Gamma \equiv \Gamma(\pspp \ra f)/\Gamma(\psp \ra f). \]
Its variation as a function of $Q(f)$ is shown in Fig.~\ref{enh}.
As $Q(f) \rightarrow \infty$, the solid and dashed contours
converge into the same point ($R^\prime_\Gamma \rightarrow 21$).
In such case $S$-wave state does not decay to $f$, but $D$-wave
does. Its partial width in $\psp$ and $\pspp$ decays comes solely
from the contribution of $\eta$, or the $D$-wave matrix element.

\subsection{\label{set3C}Final states with $Q(f)=1$}

These final states observe the 12\% rule in $\jpsi$ and $\psp$
decays. In this case, apparently one solution is identical to the
pure electromagnetic decays, with 
$\eta=5R^{\prime\prime}_{1D}(0)/(2\sqrt{2}m_c^2)$
and $R_\Gamma = 0.048$. As in the leptonic decays, their partial
widths in $\pspp$ decays are small relative to the partial widths
in $\jpsi$ decays. However, there are also other solutions with
overwhelming contribution from $\eta$ which lead to very large
partial widths in $\pspp$ decays. This can be seen from
Fig.~\ref{enhsup} at the point with $Q(f)=1$. We get the maximum
of $R_\Gamma$ of 34.0 which corresponds to another real and
positive value of $\eta$. If $\eta$ has a non-zero phase, then
$0.048 < R_\Gamma < 34.0$.

\subsection{Numerical results}

From Fig.~\ref{enhsup}, we see that except in the range
$0.52<Q(f)<2.06$ and a small range of the phase, $R_\Gamma$ is
always greater than 1. This range excludes virtually all known
decay modes except $\BBb$ which has $Q(\BBb) \approx 1$. Even
inside this range, there are other solutions by which $R_\Gamma$
(or $R^\prime_\Gamma$) is at ${\cal O}(10)$. It means
in this scenario, contrary to the na\"{\i}ve guess, the charmless
decay width of $\pspp$ is greater than that of $\jpsi$ or $\psp$.
More surprising is that $R_\Gamma$ could be as large as a few tens
for the final states with $Q(f)>1$. In general, the final states
which are enhanced in $\psp$ decays possibly have a large combined
partial width in $\pspp$ decays, especially if the phase of $\eta$
is zero or very small.

There are reasons to assume that the phase $\phi$ which is between
the matrix elements $\langle f |2S\rangle$ and $\langle f
|1D\rangle$ should be small~\cite{psippkskl}. For the decay mode
like $\rhopi$, since there is almost complete cancellation between
$\cos\theta R_{2S}(0)$ and $\eta \sin\theta$ so that
$\langle \rhopi | \psp \rangle = \cos\theta
R_{2S}(0)-\eta \sin\theta \approx 0$, the phase of $\eta$
must be small. If this is to be extrapolated to all final states,
the physics solution will follow the solid contour of
Fig.~\ref{enhsup}. Another argument comes from the universal phase
between the strong and electromagnetic amplitudes of the
charmonium decays. It has been known that in the two-body decays
of $\jpsi$, the phase between the strong and electromagnetic
amplitudes is universally around $90^\circ$~\cite{kopke,suzuki}.
Recently, it has been found that this phase is also consistent
with the experimental data of $\psp$ and $\pspp$
decays~\cite{wympspp,wymphase}. Since there is no extra phase
between $2S$ and $1D$ matrix elements due to electromagnetic
interaction, as in the calculations of the leptonic decay rates of
$\psp$ and $\pspp$, the universal phase between the strong and
electromagnetic interactions implies there is no extra phase between
the two matrix elements due to the strong interaction too, $i.e.$
$\phi \approx 0$. This conclusion means, for the modes which are 
enhanced in $\psp$ decays, their partial widths in $\pspp$ decay must 
be greater than those in $\jpsi$ or $\psp$ decays by more than an order of magnitude.

In Sect.~\ref{II}, we estimate that $\BR(\jpsi \ra ggg) \simeq
(69.2\pm0.6)\%$ while $\BR(\psp \ra ggg) \simeq (18.0\pm 2.4)\%$.
Among the final states, we know that VP and VT modes have $Q(f) <
1$. For them,
$$ \begin{array}{rcl}
 \sum \BR(\jpsi \ra \mbox{VP, VT}) &\approx&  4.6\%~, \\
 \sum \BR(\psp \ra \mbox{VP, VT}) &\approx&  8.6 \times 10^{-4}~.
  \end{array} $$
Furthermore, there are final states with $Q(f) \approx 1$ 
(such as $\BBb$), for them,
$$ \begin{array}{rcl}
 \sum \BR(\jpsi \ra B\overline{B}) &\approx&  0.9\%~, \\
 \sum \BR(\psp \ra B\overline{B}) &\approx&  9.3 \times 10^{-4}~.
  \end{array} $$
After subtracting the final states which are known to have $Q(f)
<1$ and $Q(f) \approx 1$, the remaining 63.8\% of $\jpsi$ decay
with a total width of $\Gamma(\jpsi \ra r.f.s.) \approx 58.1$~keV, 
and 17.8\% of $\psp$ decays with a
total width of $\Gamma(\psp \ra r.f.s.) \approx 50.1$~keV 
which are gluonic either has $Q(f)> 1$ or
$Q(f)$ unknown. Here $r.f.s.$ stands for the remaining final states.   
On the average, these final states have
$$  Q(r.f.s.) \approx 2.19~. $$
This is roughly comparable to $Q(\kskl)=2.26$.

If there is no extra phase between $2S$ and $1D$ matrix elements
as argued above, then $R_{\Gamma}$ takes the maximum possible 
value with $R_{\Gamma} \approx 51.6$ for $Q(r.f.s.)\approx 2.19$. 
With this value, we find $\Gamma(\pspp \ra r.f.s.) = 
R_{\Gamma} \times \Gamma(\jpsi \ra r.f.s.) \approx 3.0$~MeV for the
%\approx 51.6 \times 58.1~\hbox{keV} \approx 3.0$~MeV for the 
partial width of these remaining final states in $\pspp$ decays, which is 13\% of the total $\pspp$ width.

The above calculation takes the averaged $Q(f)$ for the final states with $Q(f)>1$ and $Q(f)$ unknown, so it merely serves as a rough estimation. 
The exact value of the partial width should be the sum of the individual 
final states which in general have various $Q(f)$ values. At present a 
major impediment to do the accurate evaluation is the lack of experimental information. Nevertheless, if we take 13\% as charmless decays 
(the calculations done channel by channel for VP, VT and $\BBb$ modes in Table~\ref{qval} give a summed maximum possible width of 93 keV in 
$\pspp$ decays, or 0.4\% of the $\pspp$ total width) together with the 
charmonium transition contributions of 3\%~\cite{rosner} (2.5\% for 
radiative transition and 0.4\% for $\pi\pi\jpsi$), we obtain a maximum 
of $\pspp$ non-$\ddb$ decay branching fraction of 16\% in the $2S$-$1D$ mixing scenario to be compared with 18\% as summarized in 
Ref.~\cite{rosner}.

\section{Experimental Measurements}

With the expected data of a few fb$^{-1}$ at $\pspp$ from the
running CLEOc and more from the future BES-III, it is of great
importance to search for the charmless decays of $\pspp$ experimentally. 
These measurements test the $2S$-$1D$ mixing scenario on the explanation 
of the $\rhopi$ puzzle in $\jpsi$ and $\psp$ decays, and provide 
information on the relative phase between the $2S$ and $1D$ matrix 
elements, which is related to the charmonium decay dynamics.

Besides the theoretical interests, such measurements are important for 
the experiment itself. So far, in the fitting of the $\pspp$ resonance 
parameters with the scanned cross sections, it has been assumed that 
$\pspp$ decays completely into $\ddb$~\cite{lgwpp,delcopp,mrkpp,cbalpp}. 
If the non-$\ddb$ branching fraction is substantial, it must be 
considered in the fitting in order to get self-consistent results.

\subsection{Inclusive method}

A substantial non-$\ddb$ decays of $\pspp$, including charmless decays 
and charmonium transitions, originally caught attention from the 
comparison between the cross sections of the inclusive hadrons and $\ddb$ at the $\pspp$ peak. However, this is not unambiguous due to the poor 
statistics of the data  samples and the complexity of the analysis. 
In addition, results from different experiments are consistent with each 
other only marginally.

\begin{figure}[htbp]
\includegraphics[width=7cm]{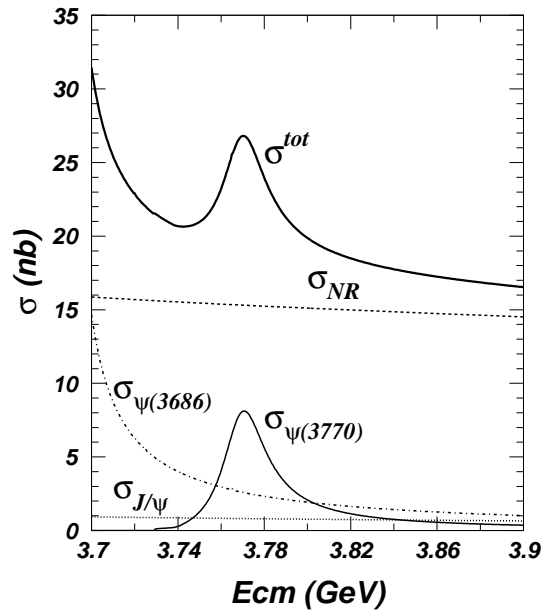}
\caption{\label{expcur}The cross section $\sigma({\EE} \ra
\mbox{hadrons})$ in the vicinity of the $\pspp$ resonance calculated 
with parameters provided by PDG. The total cross section $\sigma^{tot}$ is divided into four parts: the non-resonance $\sigma_{NR}$, the 
radiative tails of $\jpsi$ ($\sigma_{\jpsi}$) and $\psp$ 
($\sigma_{\psp}$), and the $\pspp$ resonance ($\sigma_{\pspp}$).}
\end{figure}

Fig.~\ref{expcur} shows $\sigma({\EE} \ra \mbox{hadrons})$ in the
vicinity of the $\pspp$ resonance calculated with parameters by
PDG~\cite{pdg}. The total cross section $\sigma^{tot}$
can be expressed as \beq
\sigma^{tot}=\sigma_{NR}+\sigma_{\jpsi}+\sigma_{\psp}
+\sigma_{\pspp}, \label{sigtt} \eeq which contains four parts: the
non-resonance cross section $\sigma_{NR}$, the radiative tails of
$\jpsi$ ($\sigma_{\jpsi}$) and $\psp$ ($\sigma_{\psp}$), and the
$\pspp$ resonance ($\sigma_{\pspp}$). The non-resonance cross
section is usually expressed in terms of R value and the $\mu$
pair cross section as $\sigma_{NR}=R \cdot \sigma(\EE \ra \MM)$.
The Breit-Wigner formula is adopted to depict resonances of
$\jpsi$, $\psp$ and $\pspp$, where the total decay width of $\pspp$ is 
energy dependent:
%independent for narrow resonance like $\jpsi$ or $\psp$, but
%energy dependent for $\pspp$:
$$\sigma_{\pspp}(E_{cm}) =\frac{12\pi\Gamma_{ee}\Gamma_{\pspp}(E_{cm})}
{(E^2_{cm}-M^2_{\pspp})^2+ \Gamma^2_{\pspp}(E_{cm}) M^2_{\pspp}}~,$$
with
\beq \Gamma_{\pspp} (E_{cm})
= C_{\Gamma} \left[\frac{p^3_{D^0}}{1+(r p_{\scriptstyle D^0})^2}
      + \frac{p^3_{D^{\pm}}}{1+(r p_{\scriptstyle D^{\pm}})^2}\right],
\label{gampp} \eeq
where $p$ is the $D^0$ or $D^{\pm}$ momentum, $r$ is the classical
interaction radius, and $C_{\Gamma}$ is defined as follows:
$$ C_{\Gamma}\equiv \frac{\Gamma_{\pspp} (M_{\pspp})}
{\displaystyle \left[\frac{p^3_{D^0}}{1+(r p_{\scriptstyle D^0})^2}
      + \frac{p^3_{D^{\pm}}}{1+(r p_{\scriptstyle D^{\pm}})^2}\right]
\Bigg|_{ E_{cm}=M_{\pspp} } }.     $$ Here $\Gamma_{\pspp}
(M_{\pspp})$ is the $\pspp$ total decay width given by PDG~\cite{pdg}.
%of $\pspp$ at the resonance peak.

In previous analyses~\cite{lgwpp,delcopp,mrkpp,cbalpp}, the observed 
inclusive hadronic cross sections were fit with the theoretical one with 
the contributing terms in Eq.~(\ref{sigtt}) to obtain the resonance 
parameters, which yield the $\pspp$ cross section at the peak. The 
comparison of this cross section with the $\ddb$ cross section measured 
by tagging $D$ mesons yields the non-$\ddb$ decay branching fraction. 
An inconsistency exists in this procedure because in the fit to the 
total cross section, $\ddb$ final states was assumed to saturate the 
$\pspp$ decays. Since light hadrons have much lower thresholds than 
$\ddb$, a large non-$\ddb$ branching fraction directly affects the shape of the resonance curve, and also indirectly through the 
energy-dependent width. Taking into account the non-$\ddb$ decays, 
Eq.~(\ref{gampp}) should be revised by including another term, that is
$$
\begin{array}{l} \Gamma_{\pspp}(E_{cm}) = C_{\Gamma}^{\prime} \times \\
        \left[\frac{\displaystyle p^3_{D^0}}
 {\displaystyle 1+(r p_{\scriptstyle D^0})^2}
                                +\frac{\displaystyle p^3_{D^{\pm}}}
 {\displaystyle 1+(r p_{\scriptstyle D^{\pm}})^2} + C_{\mbox{non-}\ddb}
 \right]~, \end{array} $$
%% \label{cgampp} \eeq
where $C_{\mbox{non-}\ddb}$ is proportional to the part of the 
width from non-$\ddb$ decays, and
$$ C_{\Gamma}^{\prime}\equiv \frac{\Gamma_{\pspp} (M_{\pspp})}
{\displaystyle \left[\frac{p^3_{D^0}}{1+(r p_{\scriptstyle
D^0})^2}
      + \frac{p^3_{D^{\pm}}}{1+(r p_{\scriptstyle D^{\pm}})^2}
      + C_{\mbox{non-}\ddb} \right]
\Bigg|_{ E_{cm}=M_{\pspp} } } .     $$ 
With the $C_{\mbox{non-}\ddb}$ term in the expression for
$\Gamma_{\pspp}$, the fitting of the resonance curve to 
extract the resonance parameters must be done together with the
fitting of the $\ddb$ cross section. In this procedure, the
non-$\ddb$ decay branching fraction is extracted together with the
resonance parameters.

However, this method subtracts $\ddb$ cross section from the total 
inclusive one to get the non-$\ddb$ cross section which is only a 
fraction of the total inclusive cross section. It suffers from the large 
uncertainties of the measurements, so it is difficult to obtain a 
statistically significant result.

\subsection{Exclusive method}

The calculations in section~\ref{III} show that those final states
which are suppressed in $\psp$ decays relative to $\jpsi$, and
especially those ones enhanced in $\psp$ decays, will show up in
$\pspp$ decays with maximum possible partial widths more than an order of magnitude greater than their widths in $\jpsi$ or $\psp$ decays. So these
exclusive charmless modes should be searched for in $\pspp$ decays. 
This provides direct test of the calculations based on
the $2S$-$1D$ mixing scheme. Here we discuss three typical
exclusive modes: VP mode which is suppressed in $\psp$ decays
relative to $\jpsi$, PP mode which is enhanced, and of particular
interest, the $B\overline{B}$ and $\phi f_0(980)$ modes which
observe the 12\% rule.

To measure the exclusive VP mode in $\pspp$ decays by $\EE$
experiments, the contribution from non-resonance virtual photon
amplitude  and its interference with the resonance must be treated
with care. A recent study on the measurement of $\pspp \ra VP$ in
$\EE$ experiments shows~\cite{wympspp} that with the decay rate
predicted by the $S$- and $D$-wave mixing, the interference
between the three-gluon decay amplitude and the continuum
one-photon amplitude leads to very small cross sections for some
VP modes, e.g. $\rhopi$ and $K^{*+}K^-+c.c.$, due to the
destructive interference, but much larger cross sections for other
VP modes, e.g. $K^{*0} \overline{K}^0 + c.c.$ due to the
constructive interference. In another word, although the branching
fractions of $\rho^0 \pi^0$ and $K^{*0} \overline{K}^0$ differ by
only a fraction due to SU(3) symmetry breaking~\cite{haber}, their
production cross sections in $\EE$ collision differ by one to
two orders of magnitude.

Among the PP modes, there is the $\kskl$ final state which decays
only through strong interaction and does not couple to virtual
photon~\cite{psippkskl,haber}.  There is no complication of
electromagnetic interaction and the interference between it and
the resonance. So the observed $\kskl$ in $\EE$ experiment is
completely from resonance decays. In the $2S$-$1D$ mixing scheme,
with the BES recently reported $\kskl$ branching fractions in
$\jpsi$~\cite{jpsikskl} and $\psp$~\cite{psipkskl} decays as
inputs,
it is estimated~\cite{psippkskl} $(1.2 \pm 0.7)\times 10^{-6} <
\BR(\psi^{\prime \prime} \rightarrow  K^0_S K^0_L) < (3.8\pm
1.1)\times 10^{-5}$~\cite{change}. If there is no extra phase
between $\langle \kskl | 2^3S_1 \rangle$ and $\langle \kskl |
1^3D_1 \rangle$, then its branching fraction is at the upper bound.
With 17.7~pb$^{-1}$ $\pspp$ data, BES has set an upper
limit~\cite{bespsppkskl}, which is still beyond the sensitivity
for testing the above prediction. More precise determination
of this branching fraction is expected from the analysis based on
larger data samples of CLEOc and BES-III.

However, for other PP modes, or more generally other final states
which are enhanced in $\psp$ decays, in $\EE$ experiments there is
still the complication from the non-resonance virtual photon
amplitude and its interference with the resonance.

Of particular interest are the final states with $Q(f) \approx 1$.
These are the $B\overline{B}$ modes and the vector-scalar mode
$\phi f_0(980)$~\cite{liufeng}. As discussed in subsection~\ref{set3C} 
%%%Sect.~\ref{III}.C,
for $Q(f)=1$, there are two real and positive solutions with
$R_\Gamma=0.048$ and $R_\Gamma=34.0$. These two solutions are three
orders of magnitude apart, their branching fractions are extremely 
sensitive to the relative phase between the $2S$ and $1D$ matrix elements. The $B\overline{B}$ branching fractions in $\jpsi$ decays are at 
${\cal O} (10^{-3})$, while $\phi f_0(980)$ is $(3.2 \pm 0.9) 
\times 10^{-4}$. If the physics solution of $\eta$ is the larger one of the two real values, then the $B\overline{B}$ branching fraction in $\pspp$ decay would be at ${\cal O} (10^{-4})$  and $\phi f_0(980)$ would be $4.2 \times 10^{-5}$, which can be observed in the $\pspp$ data sample over 1~fb$^{-1}$.

\section{Summary}

Based on the available experimental information of $\jpsi$ and
$\psp$ decays, we calculate the charmless decays of $\pspp$ by
virtue of the $S$- and $D$-wave charmonia mixing scheme which was
proposed to explain the large $\pspp \ra e^+e^-$ partial width and
the $\rhopi$ puzzle. We find that this leads to a possible large
branching fraction, up to 13\%, of the charmless final state in $\pspp$
decays. Although the calculation is semi-quantitative, it demonstrates
that a large charmless branching fraction in $\pspp$ decays can
well be explained.

\acknowledgments

This work is supported in part by the 100 Talents Program of CAS
under Contract No. U-25.


\begin{thebibliography}{**}
\bibitem{resvp} These are the simple weighted average of the BES
results~\cite{besrop,besksk,besogp} and the CLEOc
results~\cite{cleocvp}, regardless the different treatment of the
continuum amplitudes. %%The $\BR(K^+ \overline{K}^* (892)^- +c.c.)$
\end{thebibliography}
\end{document}